\documentclass[11pt]{article}
\usepackage{graphicx}
\usepackage{amsmath}
\usepackage{amssymb}

\newfont{\ensmathquatorze}{msbm10 scaled 1400}
\newfont{\ensmathonze}{msbm10 scaled 1100}
\newfont{\ensmathdix}{msbm10}
\newfont{\ensmathneuf}{msbm10 scaled 833}
\newfont{\ensmathhuit}{msbm10 scaled 694}
\newfam\ensmathfam
\textfont\ensmathfam=\ensmathonze
\scriptfont\ensmathfam=\ensmathdix
\scriptscriptfont\ensmathfam=\ensmathhuit

\def\be{\begin{equation}}
\def\ee{\end{equation}}

\def\bea{\begin{eqnarray}}
\def\eea{\end{eqnarray}}
\def\beann{\begin{eqnarray*}}
\def\eeann{\end{eqnarray*}}

\newcommand{\ket}[1]{|\kern.3ex#1\kern.3ex\rangle}
\newcommand{\bra}[1]{\langle\kern.3ex #1 \kern.3ex|}

\newcommand{\smean}[1]{\langle #1 \rangle} 

\newcommand{\EXP}[1]{{\mbox{\large e}}^{#1}}         

\newcommand{\re}{\mathop{\mathrm{Re}}\nolimits}      

\def\I{{\rm i}}                  
\def\D{{\rm d}}                  
\def\Dc{{\rm D}}                 

\newcommand{\drond}[2]{\frac{\partial #1}{\partial #2}} 

\newcommand\ab{{\alpha\beta}}

\begin{document}

\title{Quantum oscillations in mesoscopic rings and anomalous diffusion}


\author{Christophe Texier$^{(a,b)}$ and Gilles Montambaux$^{(b)}$}

\date{November 5, 2004}

\maketitle


{\small

\noindent
$^{(a)}$Laboratoire de Physique Th\'eorique et Mod\`eles Statistiques, 
associ\'e au CNRS, 
B\^at. 100,\\
$^{(b)}$Laboratoire de Physique des Solides, associ\'e au CNRS, 
B\^at. 510,\\
$\phantom{^{(b)}}$Universit\'e Paris-Sud, F-91405 Orsay cedex, France.

}

\begin{abstract}
We consider the weak localization correction to the conductance of
a ring connected to a network. We analyze the harmonics content of
the Al'tshuler-Aronov-Spivak (AAS) oscillations and we show that the
presence of  wires connected to the ring is responsible for a
behaviour different from the one predicted by AAS. 
The physical origin of this behaviour is the anomalous diffusion
of Brownian trajectories around the ring, due to the diffusion in the wires.
We show that this problem is related to the anomalous diffusion along
the skeleton of a comb.
We study in detail the winding properties of Brownian curves around 
a ring connected to an arbitrary network.
Our analysis is based on the spectral determinant and on the introduction of
an effective perimeter probing the different time scales. A general expression
of this length is derived for arbitrary networks. More specifically we
consider the case of a ring connected to wires, to a square network, and to a
Bethe lattice.
\end{abstract}

\noindent
PACS numbers~: 73.23.-b~; 73.20.Fz~; 72.15.Rn~; 02.50.-r~; 05.40.Jc.






\section{Introduction}

At a classical level, a network made of diffusive wires can be
described as an ensemble of classical resistances. Quantum
corrections bring a small sample specific contribution
whose disorder average is called the weak localization (WL) 
correction. 
This contribution is sensitive to a magnetic field and makes the average 
conductivity of a phase coherent ring be a periodic function of the 
magnetic flux $\phi$ through the ring with periodicity $\phi_0/2$, 
where $\phi_0=h/e$ is the flux quantum. 
This is the so-called Al'tshuler-Aronov-Spivak (AAS) effect 
\cite{AltAroSpi81}. 
The case of an isolated ring was considered by AAS who showed that the 
average correction to the classical conductivity varies as~: 
\be\label{Deltasig}
\Delta\sigma(\theta) = - \frac{e^2}{h}L_\varphi
\frac{\sinh(L/L_\varphi)}{\cosh(L/L_\varphi)-\cos(\theta)}
\ee 
where $L$ is the perimeter and $\theta=4\pi\phi/\phi_0$ the reduced flux.
Phase breaking mechanisms are taken into account through the characteristic
length $L_\varphi$ called the phase coherence length.
The harmonics of the oscillations
\be\label{AAS}
\Delta\sigma^{(n)}=
\int_{0}^{2\pi}\frac{\D\theta}{2\pi}
\Delta\sigma(\theta)\:\EXP{-\I n\theta} 
= - \frac{e^2}{h}L_\varphi\:\EXP{- |n| L/L_\varphi} 
\ee
decay exponentially with the perimeter $L$ of the ring and the
order $n$ of the harmonic\footnote{ 
Note that
$\Delta\sigma^{(n)}=\Delta\sigma^{(-n)}$ is a general property related
to the symmetry $\Delta\sigma(\theta)=\Delta\sigma(-\theta)$ and
we will simply omit the absolute value in the following.
}.
The relation (\ref{AAS}) was derived for an isolated ring and it is not
clear, when studying the transport through a ring connected by wires to
reservoirs, how this correction to the conductivity is related to the 
correction to the conductance.
Moreover the presence of the connecting wires can seriously
modify the harmonics content of the AAS oscillations. In this
paper we show that, if the perimeter $L$ of the  ring is much
smaller than $L_\varphi$ and if the connecting wires are much
longer than $L_\varphi$, the $n$th harmonic decreases like 
$\exp-n \sqrt{N_aL/L_\varphi}$, that is faster than (\ref{AAS}). $N_a$ is the 
number of wires attached to the ring.

The paper is organized as follows~:
in section \ref{sec:qtr} we briefly recall the general expression of the 
WL correction on a network and we present some results for the conductance 
through a connected ring.
The rest of the paper is devoted to the analysis of the new behaviour
of the harmonics. This is achieved by studying in detail the winding
properties of Brownian curves around a ring connected to a network.
In section \ref{sec:gf} we present our general method, based on the spectral 
determinant which encodes the necessary information on the network. 
In section \ref{sec:rctw} we consider specifically the question of a 
ring connected to one or several wires. We show that the new behaviour 
$\exp-n \sqrt{N_aL/L_\varphi}$ is the signature of an anomalous diffusion 
around the ring. 
In section \ref{sec:rctan}  we study the case where the ring is connected 
to an arbitrary network. We relate the winding properties around the ring
to the recurrent character of the Brownian motion in the network.

\begin{figure}[!ht]
\begin{center}
\includegraphics[scale=1]{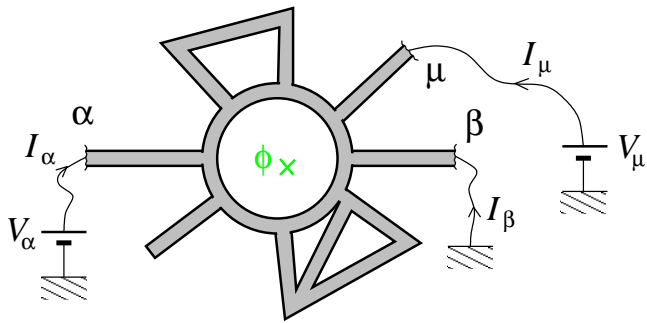}
\end{center}
\caption{\it A network connected to reservoirs and pierced by a magnetic flux
         $\phi$. 
         The wavy lines represent connection to large contacts (external 
	 reservoirs). \label{fig:graph1d}}
\end{figure}


\section{Quantum transport through a connected ring\label{sec:qtr}}

Recently, we have obtained a general expression for the
weak localization (WL) correction on a network
\cite{TexMon04,TexMonAkk05}. Let us remind first how it reads for
a wire. The classical conductance of a wire of section $s$ and length
$L$ is given by the Ohm's law $G^{\rm cl}=\sigma_0 s/L$ where $\sigma_0$ 
is the Drude conductivity. This result can be rewritten in terms of the 
total transmission through the wire, {\it i.e.} the dimensionless 
conductance, $T^{\rm cl}_{\rm wire}= \alpha_d N_c \ell_e/L$, where $N_c$ is
the number of channels, $\ell_e$ the elastic mean free path and
$\alpha_d$ a numerical constant depending on the dimension
($\alpha_1=2$, $\alpha_2=\pi/2$ and $\alpha_3=4/3$).
  The interferences between reversed trajectories are described by
the so-called cooperon $P_c(x,x)$ which measures the return
probability for a diffusive particle and which is solution of a
diffusion equation that we will recall in section \ref{sec:gf}. These
interferences give rise to the WL correction which
is expressed as
\be\label{RES31intro} 
\Delta T_{\rm wire} = -\frac{2}{L^2} 
\int_0^L \D x\, P_c(x,x) 
\:.\ee
More generally, for a network attached to
leads $\alpha$ and $\beta$ as shown on figure~\ref{fig:graph1d},
the classical transmission coefficient is obtained by classical
Kirchhoff laws and it has the form
$T^{\rm cl}_{\alpha\beta}=\alpha_d N_c\ell_e/{\cal L}_{\alpha\beta}$ where 
the equivalent length ${\cal L}_{\alpha\beta}$ is function of the lengths 
of the wires, the topology of the network and the way it is connected to 
external reservoirs. Note that ${\cal L}_{\alpha\beta}$ is simply 
proportional to the equivalent resistance.

On such a network, because of the absence of translation
invariance, we have shown recently that the cooperon must be properly
weighted when integrated over the wires of the networks to get the
WL correction to the transmission coefficient. Equation
(\ref{RES31intro}) generalizes as \cite{TexMon04}~: 
\be\label{RES3intro} 
\Delta T_{\alpha\beta} = \frac{2}{\alpha_d N_c\ell_e} \sum_{(\mu\nu)}
\frac{\partial T^{\rm cl}_{\alpha\beta}}{\partial\,l_{\mu\nu}}
\int_{(\mu\nu)}\D x\, P_c(x,x) 
\:.\ee 
The sum runs over all wires
$(\mu\nu)$ of the network. The cooperon is integrated over each
wire $(\mu\nu)$, with a weight which is simply   the derivative of
the classical transmission with respect to its length
$l_{\mu\nu}$. This WL correction depends on the lengths of
the wires, the phase coherence length $L_\varphi$, the magnetic
field and the topology of the network.

\begin{figure}[!ht]
\begin{center}
\includegraphics[scale=1]{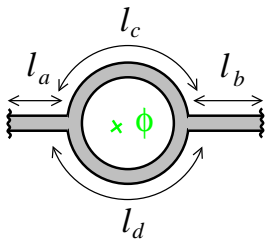}
\end{center}
\caption{\it A mesoscopic ring connected at two reservoirs.\label{fig:loop2}}
\end{figure}

We consider the transport through the ring of
figure~\ref{fig:loop2}. The classical transmission reads 
$T^{\rm cl}_{\rm ring} = \frac{\alpha_d N_c\ell_e}{l_a+l_{c/\!/d}+l_b}$, 
where $l_{c/\!/d}^{-1}=l_{c}^{-1}+l_{d}^{-1}$. To simplify the
calculations we consider below the symmetric case $l_a=l_b$ and
$l_c=l_d=L/2$. The calculation of the WL correction 
$\Delta T_{\rm ring}(\theta)$, given by (\ref{RES3intro}), requires the
construction of the cooperon in the network, explained in
\cite{TexMon04,TexMonAkk05}. 
We do not give further details nor the full result
\cite{TexMonAkk05} and we only present two limiting cases for the
Fourier decomposition $\Delta T_{\rm ring}^{(n)}~$:

\vspace{0.25cm}

\noindent
$\bullet$ {\it The weakly coherent network $L_\varphi\ll L,l_a$.}
\be\label{harmT14a} 
\Delta T_{\rm ring}^{(n)} 
\simeq - \frac{L\,L_\varphi}{4(2l_a+L/4)^2}
\left(\frac{2}{3}\right)^{2n}
\EXP{-n\,L/L_\varphi} \hspace{0.5cm}\mbox{ for } n>0 
\:.\ee 
We get
an exponential behaviour coinciding with the one predicted by AAS.
The factor $\left({2}/{3}\right)^{2n}$  is related to the probability 
to cross $2n$ times a vertex of coordinence 3 \cite{AkkComDesMonTex00}. 
This result shows that the connecting wires play an important role 
and that the determination of $L_\varphi$ from ratio of harmonics
using formula (\ref{AAS}) can lead to a wrong estimate.

\vspace{0.25cm}

\noindent $\bullet$ {\it The mesoscopic ring connected to long
wires $L\ll L_\varphi\ll l_a$.}
\be\label{harmT14b} \Delta T_{\rm ring}^{(n)} 
\simeq -
\left(\frac{L_\varphi}{2l_a}\right)^{2}
\sqrt{\frac{L}{2L_\varphi}} \:\EXP{ -n\,\sqrt{2L/L_\varphi} }
 \hspace{0.5cm}\mbox{ for } n>0 
\:.\ee 
The harmonics decay presents a strikingly different
behaviour from the one given by  AAS. This can be understood in
the following way~: 
because $L\ll L_\varphi$, whenever a diffusive trajectory encircles the
ring, it explores a distance in the arm which can be much longer
than the perimeter of the ring. This leads to an effective
perimeter $ L_{\rm eff}\simeq\sqrt{2L_\varphi L}\gg L$ which is
responsible for a decrease of the harmonics with $n$ faster than
(\ref{AAS}). In the next sections, we examine in detail
the effect of the arms on the winding properties around the loop
in order to analyze the origin of the new behaviour (\ref{harmT14b}).


\section{Winding of Brownian trajectories and spectral determinant
         \label{sec:gf}}

In this section, we recall how the harmonics of the WL correction,
$\Delta T^{(n)}_{\rm ring}$, can be related to the winding properties
of diffusive trajectories around the ring.
For this purpose we introduce the probability to start from a point $x$ and 
come back to it after a time $t$ having performed $n$ windings around the 
loop. This probability can be written with a path integral as~:
\be 
W_n(x,x;t) 
= \int_{x(0)=x}^{x(t)=x}{\cal D}x(\tau)
\:\EXP{-\frac14\int_0^t\D\tau\,\dot x(\tau)^2} 
\delta_{n,{\cal N}[x(\tau)]} 
\ee
where ${\cal N}[x(\tau)]=\int_0^t\D\tau\,\dot x(\tau)$ is the winding 
number of the diffusive trajectory $x(\tau)$. 
We have chosen a diffusion constant equal to unity $D=1$.
If we write 
$\delta_{n,{\cal N}}=\int_{0}^{2\pi}\frac{\D\theta}{2\pi} 
\EXP{\I ({\cal N}-n)\theta}$, 
the coupling of the winding number to the conjugate
parameter $\theta$ in the action is interpreted as the coupling to
a magnetic flux. The Laplace transform with respect to time
relates the probability $W_n(x,x;t)$ to the cooperon $P_c(x,x)$~:
\bea
\int_0^\infty\D t\,W_n(x,x;t)\,\EXP{-\gamma t}
&=& \int_0^{2\pi}\frac{\D\theta}{2\pi}\EXP{-\I n\theta}
\int_0^\infty\D t\,\EXP{-\gamma t}\int_{x(0)=x}^{x(t)=x}{\cal D}x\:
\EXP{-\frac14\int_0^t\D\tau\,\dot
     x^2+\I\frac{\theta}{L}\int_0^t\D\tau\,\dot x} \\\label{Wn} 
&=&
\int_{0}^{2\pi}\frac{\D\theta}{2\pi}\:\EXP{-\I n\theta}\:P_c(x,x)
\:.\eea
The cooperon $P_c(x,x')$ is solution of a diffusion-like equation 
$[\gamma - \Dc_x^2]P_c(x,x')=\delta(x-x')$ where the covariant derivative 
is $\Dc_x=\frac{\D}{\D x}-\I A(x)$ with $A(x)=\theta/L$ for $x$ inside 
the ring.
In the WL theory, the cooperon describes the contribution of quantum 
interferences, that are limited by phase breaking mechanisms. 
In this respect, the Laplace parameter $\gamma$ plays the role of a 
phenomenological parameter that selects the diffusive trajectories for times 
$t<1/\gamma$. 
This parameter is related to the phase coherence length $L_\varphi$, that 
gives the length scale over which quantum interferences can occur, 
by $\gamma=1/L_\varphi^2$.
The $n$-th harmonics of the WL correction $\Delta T_{\rm ring}^{(n)}$
is given by an integral over $x$ of the Laplace transform 
$\int_0^\infty\D t\,W_n(x,x;t)\,\EXP{-\gamma t}$. The integration over
the network is performed by weighting the wires with the coefficients
given in eq.~(\ref{RES3intro}).

The relation between WL and properties of the Brownian motion was 
formulated in other works like \cite{ComDesOuv90,Mon95}.

\subsection{Probability averaged over the network}

First, we do not consider the dependence on the initial point
$x$ and average over it. We define $W_n(t)=\int\D x\,W_n(x,x;t)$.
In order to study this quantity it is useful to introduce the spectral 
determinant, defined as $S(\gamma)=\prod_n(\gamma+E_n(\theta))$, where
$\{E_n(\theta)\}$ is the spectrum of the operator $-\Dc_x^2$. The spectral
determinant is related to the cooperon in the following way~:
$
\int\D x\,P_c(x,x)={\rm Tr}\{\frac1{\gamma-\Dc_x^2}\}=\sum_n\frac1{\gamma+E_n}
=\frac{\partial}{\partial\gamma}\ln S(\gamma)
$. 
Therefore the probability $W_n(t)$ can be conveniently written as~:
\be\label{relaWS} 
\int_0^\infty\D t\,W_n(t)\,\EXP{-\gamma t}
=\int_{0}^{2\pi}\frac{\D\theta}{2\pi}\:\EXP{-\I n\theta}
\drond{}{\gamma}\ln S(\gamma) 
\:.\ee
The interest to introduce the spectral determinant is that
it is a global quantity encoding all informations about the
spectrum.
This approach is especially efficient for networks thanks to the 
compact expression of $S(\gamma)$  obtained in \cite{PasMon99}~: 
it can be related to the determinant
of a finite size matrix that encodes the information on the topology of the
networks, the lengths of the wires, the magnetic fluxes and the way the 
network is connected to external reservoirs. This relation is briefly
recalled in the appendix~\ref{sec:sd}.

\subsection{Probability for a fixed initial condition\label{sec:fic}}

It is also interesting to study the probability $W_n(x_0,x_0;t)$ for $n$ 
windings in a time $t$ when the initial condition $x_0$ is fixed.
This requires some local information which is obtained from 
the construction of $P_c(x_0,x_0)$ inside the network (the 
expression can be found in \cite{TexMonAkk05}). 
We propose here a method to obtain this local information still
using the spectral determinant. This latter encodes a global information
since it results from a spatial integration of the cooperon over the 
network.
It is related to a sum over the eigenvalues of the Laplace operator 
$\sum_n\frac1{\gamma+E_n}$. On the other hand $P_c(x_0,x_0)$ requires 
some local information on the eigenfunctions.
In order to extract this local information we introduce the modified 
cooperon $P_c^{(\lambda_0)}$, solution of 
$
[\gamma - \Dc_x^2+\lambda_0\delta(x-x_0)]P_c^{(\lambda_0)}(x,x')
=\delta(x-x') 
$. 
The corresponding spectral determinant  is
$
\partial_\gamma\ln S^{(\lambda_{0})}(\gamma)=\int\D x\,P_c^{(\lambda_0)}(x,x)
$.
Its expansion in powers of the parameter $\lambda_0$ leads to~:
$
\partial_\gamma\ln S^{(\lambda_0)}(\gamma) =
\partial_\gamma\ln S(\gamma) + \lambda_0\partial_\gamma P_c(x_0,x_0)
+O(\lambda_0^2)
$.
The cooperon at $x_0$ can be obtained by computing the spectral determinant
$S^{(\lambda_0)}(\gamma)$ for a $\delta$ added at $x_0$~:
\be\label{localcoop}
P_c(x_0,x_0)=
\frac{\D}{\D\lambda_0}\ln S^{(\lambda_0)}(\gamma)\Big|_{\lambda_0=0}
\:,\ee
and inserted in (\ref{Wn}).
For a network, the computation of $S^{(\lambda_0)}(\gamma)$ is especially
simple~:
$\lambda_0$ is interpreted as the parameter involved in 
the boundary condition at a vertex at $x_0$ (see appendix~\ref{sec:sd}).


\section{Winding around a ring with an arm\label{sec:rctw}}

In this section we study how the winding properties inside a ring are 
affected by the presence of an arm.
We first recall the simple case of the isolated ring and  consider
the case of the ring with one arm (figure~\ref{fig:ouvrebouteille}a) for 
different kinds of boundary conditions at the end of the arm.

\subsection{Spectral determinant and winding properties}

\noindent{\bf A reminder~: the isolated ring.--}
In order to illustrate our formalism, we start by recalling few simple
results about the isolated ring, that will be useful for the following.
The quantity at the basis of our analysis is the spectral determinant,
that reads for the isolated ring~:
$S(\gamma)=2(\cosh\sqrt{\gamma}L-\cos\theta)$.
The relation (\ref{relaWS}) shows that the Laplace transform of the 
probability $W_n(t)$ require a Fourier transform of 
$
\frac{\partial}{\partial\gamma}\ln S(\gamma)
=\frac{L}{2\sqrt\gamma}
\frac{\sinh\sqrt{\gamma}L}{\cosh\sqrt{\gamma}L-\cos\theta}
$.
We obtain 
\be
\int_0^\infty\D t\, W_n(t)\,\EXP{-\gamma t} 
= \frac{L}{2\sqrt\gamma}\EXP{-n\sqrt\gamma L}
\:,\ee
which is the exponential behaviour recalled in eq.~(\ref{AAS}).
It is worth noticing that the translation invariance of the system implies 
that $W_n(t)=LW_n(x,x;t)$.
The inverse Laplace transform leads to 
\be\label{normaldiff}
W_n(x,x;t) = \frac1{2\sqrt{\pi t}}\EXP{-\frac{(nL)^2}{4t}}
\ee
which characterizes normal diffusion in the ring. 
We can study the scaling of the winding number with time by computing
$n_t=\sqrt{\smean{n^2}_t}$, where $\smean{\cdots}_t$ denotes averaging
over all trajectories for a time $t$. 
The ring possesses a characteristic time scale~: the Thouless time.
In units such that $D=1$, it reads $\tau_L=L^2$. It is the time needed by a 
diffusive trajectory to explore the ring. 
In the short time limit $t\ll\tau_L$, we obtain 
$n_t\simeq\sqrt2\exp-\frac{L^2}{8t}$ 
and in the long time limit $t\gg\tau_L$ we get $n_t\simeq\sqrt2\,t^{1/2}/L$.

\vspace{0.25cm}

\begin{figure}[!ht]
\begin{center}
\includegraphics[scale=1]{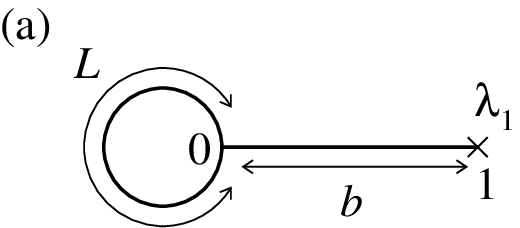}
\hspace{1cm}
\includegraphics[scale=0.75]{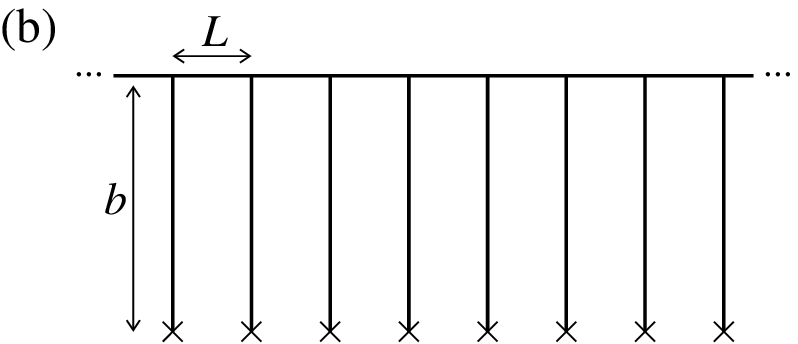}
\end{center}
\caption{Left~:
         {\it A network with one loop and one arm.
         A general boundary condition ($\lambda_1$) is chosen at 
         vertex 1. $\lambda_1=0$ describes a reflecting boundary (isolated 
         network).
         $\lambda_1=\infty$ descibes an absorbing boundary (connection to 
         a reservoir).}
         Right~: 
         {\it The study of the winding in the ring is
         equivalent to study the displacement along the skeleton of a
         comb.}
         \label{fig:ouvrebouteille}}
\end{figure}

\noindent{\bf Dirichlet boundary at the end of the arm.--}
We now consider the winding in a ring with an arm.
A Dirichlet boundary at the end of the wire describes absorption 
(connection to a reservoir). 
In terms of the parameter introduced in the  appendix~\ref{sec:sd}, 
this corresponds to $\lambda_1=\infty$.
The spectral determinant of the ring of figure~\ref{fig:ouvrebouteille}a 
is easily obtained (see \cite{PasMon99,AkkComDesMonTex00,TexMonAkk05} 
and  appendix~\ref{sec:sd}). 
It presents a structure similar to the one of the isolated ring,
\be \label{SarmDir}
\sqrt\gamma S_{\rm Dir}(\gamma) 
=2\sinh\sqrt{\gamma}b \,
\left[\cosh\sqrt{\gamma} L_{\rm eff}^{\rm Dir}-\cos\theta\right] 
\:,\ee
where we have introduced $ L_{\rm eff}^{\rm Dir}$, defined by 
\be\label{leffDir}
\cosh\sqrt{\gamma} L_{\rm eff}^{\rm Dir}=\cosh\sqrt{\gamma}L+
\frac12\sinh\sqrt{\gamma}L\,\coth\sqrt{\gamma}b
\:.\ee
Interestingly, the flux dependance 
$S(\gamma)\propto\cosh\sqrt{\gamma} L_{\rm eff}-\cos\theta$
immediatly leads to the behaviour 
$\int_0^\infty\D t\, W_n(t)\,\EXP{-\gamma t}
\propto\EXP{-n\sqrt\gamma  L_{\rm eff}}$.
In general the length $ L_{\rm eff}$ depends on $\gamma$. Since 
$\gamma$ is the parameter conjugate to the time $t$, it probes times 
$t\sim1/\gamma$. Therefore $ L_{\rm eff}(\gamma)$ is the 
effective perimeter for winding trajectories at a time scale $t=1/\gamma$.

In the limit $b\to0$ the spectral determinant becomes
$S_{\rm Dir}(\gamma)|_{b=0}=\sinh\sqrt{\gamma}L/\sqrt{\gamma}$ which 
is the result for a wire with Dirichlet boundary at its ends.
The reservoir acting as a phase breaker, the phase sensitivity is lost 
since the reservoir is now located in the ring.

\vspace{0.25cm}

\noindent{\bf Neumann boundary at the end of the arm.--}
It is also interesting to study the spectral determinant
for Neumann boundary at the end of the arm, $\lambda_1=0$, which describes 
a reflection of the diffusive particle (isolated network). 
In this case \cite{AkkComDesMonTex00}
\be
S_{\rm Neu}(\gamma) =2\cosh\sqrt{\gamma}b \,
\left[\cosh\sqrt{\gamma} L_{\rm eff}^{\rm Neu}-\cos\theta\right]
\ee
where the effective perimeter now reads
\be\label{leffNeu}
\cosh\sqrt{\gamma} L_{\rm eff}^{\rm Neu}=\cosh\sqrt{\gamma}L+
\frac12\sinh\sqrt{\gamma}L\,\tanh\sqrt{\gamma}b
\:.\ee
In this case the limit $b\to0$ leads to the spectral determinant of an
isolated ring
$S_{\rm Neu}(\gamma)|_{b=0}=2(\cosh\sqrt{\gamma}L-\cos\theta)$.

\vspace{0.25cm}

We now analyze the different behaviours in a time representation to 
understand their physical significance in the diffusion problem.
To analyze the diffusion at time $t$, we have to consider $\gamma=1/t$. 
In the following we keep using the notation $\gamma=1/L_\varphi^2$ for 
convenience.
In addition to the Thouless time $\tau_L=L^2$ needed to explore the ring,
there is a second characteristic time $\tau_b=b^2$ which is the typical 
time required to explore the arm.

\vspace{0.25cm}

\noindent$\bullet$ {\bf Short times 
\mathversion{bold}$t\ll\tau_L,\tau_b$\mathversion{normal}
({\it i.e.} $L_\varphi \ll L,b$).--}
\mathversion{normal}
In this case the arm is explored over a distance smaller than the
perimeter and the presence of the arm has a small influence. 
The precise nature of the boundary condition is not felt when diffusing inside
the loop, since the trajectories encircling the loop do not have enough time to
reach the end of the arm ($t\ll\tau_b$). This is reflected by the
fact that $ L_{\rm eff}^{\rm Dir}= L_{\rm eff}^{\rm Neu}$ in this limit.
From the expressions (\ref{leffDir},\ref{leffNeu}), we obtain
the effective parameter 
$ L_{\rm eff}\simeq L+L_\varphi\ln(3/2)$. We recover the behaviour of
the isolated ring $n_t\simeq\sqrt2\exp-\frac{L^2}{8t}$.

\vspace{0.25cm}

\noindent$\bullet$ {\bf Intermediate times 
\mathversion{bold}$\tau_L\ll t\ll\tau_b$\mathversion{normal}
({\it i.e.} $L\ll L_\varphi \ll b$).--}
The particle can turn diffusively many times around the loop but cannot 
explore the arm up to its end. This is a limit of an infinitely long arm. 
Like in the previous case, the boundary condition plays no role. 
From (\ref{leffDir},\ref{leffNeu}), the effective length is 
$ L_{\rm eff}\simeq\sqrt{L}\,\gamma^{-1/4}=\sqrt{L L_\varphi}$. By using
(\ref{relaWS}) we obtain~:
\be
\label{Lapdis} 
\int_0^\infty\D t\,W_n(t)\,\EXP{-\gamma t}\simeq
\frac{ \sqrt{L} }{ 4\gamma^{3/4}}\,\EXP{-n\sqrt{L}\,\gamma^{1/4}}
\:,\ee 
which leads to the harmonic content (\ref{harmT14b})
\footnote{
  Note that only the exponential behaviour and the exponent of $\gamma$
  in the prefactor coincide in (\ref{harmT14b}) and (\ref{Lapdis}) 
  since this latter has been obtained by a uniform integration of 
  the cooperon over the network.
}. 
The inverse Laplace transform gives 
\be\label{distribution} 
W_n(t) =\theta(t)\frac{\sqrt{L}}{4\,t^{1/4}} 
\ \chi\!\left(\xi=\frac{n\sqrt{L}}{t^{1/4}}\right) 
\ee 
where 
$
\chi(\xi)=\frac{4}{\pi}
\int_0^\infty\D u\, \EXP{-u^4-\frac{1}{\sqrt{2}}\xi u }
\cos\left(\frac{1}{\sqrt{2}}\xi u+\frac\pi4\right)
=\frac{4}{\pi}\re( \EXP{\I\frac\pi4}\int_0^\infty\D
u\,\EXP{-\varphi(u)}) 
$ 
with
$\varphi(u)=u^4+u\,\xi\,\EXP{-\I\frac\pi4}$. $\theta(t)$ is the
Heaviside function. The function at the origin is
$\chi(0)=\frac{\Gamma(1/4)}{\pi\sqrt2}$ while it presents an
exponential tail~: 
\be\label{tail1} 
\chi(\xi)\simeq
\frac{4}{\sqrt{6\pi}}\: \frac1{\left(\xi/4\right)^{1/3}}\,
{\EXP{-3\,({\xi}/{4})^{4/3}}} \hspace{0.25cm}\mbox{for } \xi\gg1
\:.\ee 
As a result, the tail of the distribution behaves like 
\be\label{distribf} 
W_n(t) \sim
\frac1{t^{1/6}n^{1/3}} \exp{ -c \frac{ n^{4/3}}{ t^{1/3} } }
\:,\ee
where $c$ is a coefficient. 
We will see in section \ref{sec:fsp} that for a fixed initial condition 
$W_n(x,x;t)$ presents the same exponential behaviour with a 
different prefactor.
This expression shows that the winding number scales like 
$n_t\propto t^{1/4}$.

\vspace{0.25cm}

\noindent$\bullet$ {\bf Large times 
\mathversion{bold}$\tau_L,\tau_b\ll t$\mathversion{normal}
({\it i.e.} $L,b\ll L_\varphi$).--}
In this regime the diffusive trajectories can explore the arm until its 
end and the precise nature of boundary condition  matters.

For Neumann boundary condition, the expansion of  (\ref{leffNeu}) gives
$ L_{\rm eff}^{\rm Neu}\simeq\sqrt{L(L+b)}$.
It is interesting to point that since the arm is explored until its end,
this result can be simply obtained by replacing $L_\varphi$ by $L+b$ in 
the result $ L_{\rm eff}\simeq\sqrt{LL_\varphi}$ obtained for 
$L\ll L_\varphi\ll b$.
Since this effective perimeter does not depend on $\gamma$, it describes 
normal diffusion around the ring. 
However the diffusion constant is reduced due to the time spent in the arm 
$n_t\simeq \sqrt2\,t^{1/2}/\sqrt{L^2+bL}$.

For Dirichlet boundary, from (\ref{leffDir}), the effective length reads 
$ L_{\rm eff}^{\rm Dir}\simeq L_\varphi\sqrt{L/b}$. This reflects a behaviour
$\int_0^\infty\D t\,\EXP{-\gamma t}W_n(t)\propto\EXP{-n\sqrt{L/b}}$
independent on $\gamma$ which originates from the absorption at vertex $1$.
The winding number reaches a finite value at infinite time~:
$n_t\simeq\sqrt{2b/L}$.

\vspace{0.25cm}

We summarize these results in table~\ref{Latable}.

\subsection{Relation with the anomalous diffusion along the
            skeleton of a comb\label{sec:comb}} 

We discuss specifically the regime $\tau_L\ll t\ll\tau_b$ of an infinitely 
long arm. From (\ref{distribution}) we have seen that the winding number 
scales with the time like 
\be\label{subdiffusion} 
n_t\propto t^{1/4} 
\:.\ee
The prefactor can be obtained from the method explained in the 
appendix~\ref{sec:nt}.
We recall that the winding in a ring without arm (normal diffusion)
is $n_t\simeq \sqrt2\,t^{1/2}/L$.
In the presence of the arm, the Brownian trajectories can explore the arm 
over long distance when turning around the ring. For a time scale $t$ 
the effective perimeter of winding trajectories is 
$ L_{\rm eff}(\gamma=1/t)\simeq \sqrt{L}\,t^{1/4}$.
From this simple heuristic argument we recover the subdiffusive behaviour
$n_t\sim t^{1/2}/ L_{\rm eff}\sim t^{1/4}/\sqrt{L}$.
We stress that ``subdiffusion'' refers to the time dependence of the 
winding number around the flux, and not to the motion inside the wires 
which obeys normal diffusion.
 
This problem is similar to the known problem of the
diffusion along the skeleton of a comb~: the diffusion in the ring is the
periodisation of the diffusion along the skeleton of the comb (see
figure~\ref{fig:ouvrebouteille}). This problem has been studied by
a different method in refs. \cite{WeiHav86} and \cite{BalHavWei87} 
(note that the reference \cite{BalHavWei87} corrected a wrong assumption 
about  the nature of the distribution made in \cite{WeiHav86}), however the 
power law in front of the exponential was not given.
The problem of diffusion along the skeleton of a comb  belongs to a 
broader class of problems~: the diffusion of a particle along a line
with an arbitrary distribution of the waiting time $\tau$ spent on 
each site.
It was shown that if the distribution of the waiting time presents an 
algebraic tail $P_1(\tau)\propto\tau^{-1-\mu}$ with $0<\mu<1$, 
the diffusion is subdiffusive with $n_t\sim t^{\mu/2}$ \cite{BouGeo90}.
In the case of the diffusion in the comb, the distribution of the trapping 
time by the arm is given by the first return probability for the 
one-dimensional diffusion~: $P_1(\tau)\propto\tau^{-3/2}$ and we recover
$n_t\sim t^{1/4}$.

\mathversion{bold}
\subsection{The case of $N_a$ arms} 
\mathversion{normal}

By using the mapping between the diffusion inside the
ring and in the comb we immediately get the result for $N_a$
arms~: the trajectory encounters $N_a$ arms for one turn, which
corresponds to $N_a$ steps of length $L/N_a$ in the comb. Then the
result for $N_a$ arms is obtained by a substitution $n\to nN_a$
and $L\to L/N_a$ which leads to 
$\int_0^\infty\D t\,W_n(t)\,\EXP{-\gamma t}
\simeq \frac{ \sqrt{L/N_a} }{4\gamma^{3/4}}\,\EXP{-n\sqrt{N_aL}\,\gamma^{1/4}}
$. For $N_a=2$ the exponential dependence agrees with (\ref{harmT14b}).

\noindent$\bullet$ {\it $N_a$ arms attached regularly.--}
The result given by the previous simple argument can also be obtained from 
a calculation of $S(\gamma)$ for several arms. If $N_a$ arms of length $b$ 
are attached regularly around the ring we obtain 
\be
S(\gamma) = \left(\frac{2\sinh\sqrt{\gamma}b}{\sqrt{\gamma}}\right)^{N_a}
\prod_{n=1}^{N_a}
\left[
  \cosh(\frac{\sqrt{\gamma}L}{N_a}) - \cos(\frac{2\pi n+\theta}{N_a})
  +\frac12\sinh(\frac{\sqrt{\gamma}L}{N_a}) \coth\sqrt{\gamma}b
\right]
\ee
for Dirichlet boundary conditions at the end of the arms. 
For $N_a=1$ we recover (\ref{SarmDir}).
(The spectral determinant for Neumann boundary conditions is given in 
chapter 5 of \cite{AkkMon04}).
For short times, $t\ll\tau_L,\tau_b$, we get
$L_{\rm eff}\simeq L+N_aL_\varphi\ln(3/2)$. 
The term $N_aL_\varphi\ln(3/2)$ has the same origin as the term $(2/3)^{2n}$
in (\ref{harmT14a})~: it is related to the probability to cross the $N_a$
vertices of coordinence $3$ for a trajectory of winding $n=1$
\cite{AkkComDesMonTex00}. For intermediate times $\tau_L\ll t\ll\tau_b$, the
expansion of the spectral determinant shows that 
$L_{\rm eff}\simeq\sqrt{N_aLL_\varphi}$, in agreement with the above 
discussion.
For large times $\tau_L,\tau_b\ll t$, the effective length reads
$L_{\rm eff}\simeq L_\varphi\sqrt{N_aL/b}$ that reflects the absorption of 
the particle by the reservoirs after a finite time.
For $N_a=2$, that is for the ring of figure~\ref{fig:loop2},
we give the expression of the effective perimeter~:
\be\label{leff2arms}
\cosh\sqrt{\gamma}L_{\rm eff} = \cosh\sqrt{\gamma}L
+\sinh\sqrt{\gamma}L\,\coth\sqrt{\gamma}b
+\frac12\sinh^2(\sqrt{\gamma}L/2)\,\coth^2\sqrt{\gamma}b
\:.\ee

\noindent$\bullet$ {\it $N_a$ arms attached at the same point.--}
Note that the study of the effect of several arms is easier if we 
consider the case of $N_a$ arms attached at the same point. It immediatly
follows that
\be
S(\gamma)=2\left(\frac{\sinh\sqrt{\gamma}b}{\sqrt{\gamma}}\right)^{N_a}
\left(
  \cosh{\sqrt{\gamma}L} - \cos{\theta}
  +\frac{N_a}{2}\sinh{\sqrt{\gamma}L} \coth\sqrt{\gamma}b
\right)
\ee
for Dirichlet conditions, which leads to 
$\cosh\sqrt{\gamma}L_{\rm eff} = \cosh\sqrt{\gamma}L
+\frac{N_a}{2}\sinh\sqrt{\gamma}L\,\coth\sqrt{\gamma}b$.
In the short time limit, $t\ll\tau_L,\tau_b$, 
we obtain~: $L_{\rm eff}\simeq L+L_\varphi\ln((N_a+2)/2)$, whose second term
originates from the fact that each winding requires to cross one vertex
of coordinence $N_a+2$ \cite{AkkComDesMonTex00}. For large time $t\gg\tau_L$,
the winding properties are exactly similar to the case of regularly attached 
arms.

\subsection{Fixing the starting point\label{sec:fsp}} 

In this paragraph we study the distribution of
windings with a fixed starting point $x_0$. We choose the origin
of the Brownian trajectory to be at the vertex $0$ for simplicity 
(see figure~\ref{fig:ouvrebouteille}a).
Following \S\ref{sec:fic}
we consider the new spectral determinant $S^{(\lambda_0)}(\gamma)$
for generalized boundary condition with parameter $\lambda_0$ at
the vertex $0$. It is easy to obtain~: 
\be 
S^{(\lambda_0)}(\gamma) =
S(\gamma) + \lambda_0\frac{\sinh\sqrt\gamma L\sinh\sqrt\gamma b}{\gamma} 
\:.\ee 
By using (\ref{localcoop}) we immediately get
the cooperon at the vertex 
\be\label{Pc00}
P_c(0,0) =\frac{1}{2\sqrt\gamma}
\frac{\sinh\sqrt\gamma L}{\cosh\sqrt{\gamma} L_{\rm eff}-\cos\theta} 
\ee
from which 
\be\label{Wn00}
\int_0^\infty\D t\:W_n(0,0;t)\:\EXP{-\gamma t} = \frac{1}{2\sqrt\gamma}
\frac{\sinh\sqrt\gamma L}{\sinh\sqrt{\gamma} L_{\rm eff}}
\:\EXP{-n\sqrt{\gamma} L_{\rm eff}} 
\:.\ee
In the limit $L\ll L_\varphi\ll b$, it reads
\be\label{anobeh} 
\int_0^\infty\D t\:W_n(0,0;t)\:\EXP{-\gamma t} 
\simeq
\frac{\sqrt{L}}{2\gamma^{1/4}}\EXP{-n\sqrt{L}\gamma^{1/4}}
\:.\ee 
The inverse Laplace transform gives 
\be\label{Wn11}
W_n(0,0;t) = \theta(t)\frac{\sqrt{L}}{2\,t^{3/4}} \
\psi\!\left(\xi=\frac{n\sqrt{L}}{t^{1/4}}\right) 
\ee 
where 
$
\psi(\xi)
=\frac{4}{\pi}\re( \EXP{-\I\frac\pi4}
\int_0^\infty\D u\,u^2\,\EXP{-\varphi(u)})
$.
The function $\varphi(u)$ is defined above.
At the origin we have $\psi(0)=\frac{\Gamma(3/4)}{\pi\sqrt2}$
while $\psi(\xi)$ presents an exponential tail~:
\be\label{tail2}
\psi(\xi)\simeq\frac{4}{\sqrt{6\pi}}\:
\left(\xi/4\right)^{1/3}\, {\EXP{-3\,({\xi}/{4})^{4/3}}}
\hspace{0.25cm}\mbox{for } \xi\gg1
\:.\ee
The tail of $W_n(0,0;t)$ then reads
\be
W_n(0,0;t) \propto
\frac{n^{1/3}}{t^{5/6}} \exp{ -c \frac{ n^{4/3}}{ t^{1/3} } }
\:.\ee
The exponential behaviour is the same in (\ref{tail1}) and (\ref{tail2})
while the exponent of the prefactor changes because we have fixed
the initial condition instead of averaging over it.


\section{Ring connected to a network\label{sec:rctan}} 

\begin{figure}[!ht]
\begin{center}
\includegraphics[scale=0.6]{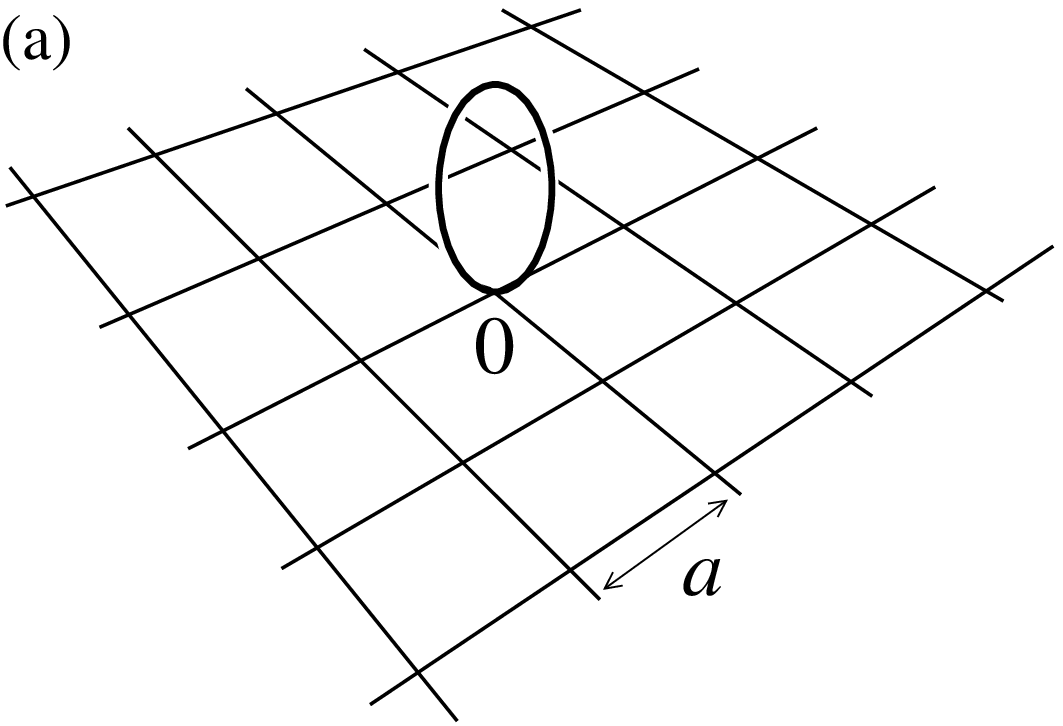}
\hspace{1cm}
\includegraphics[scale=0.6]{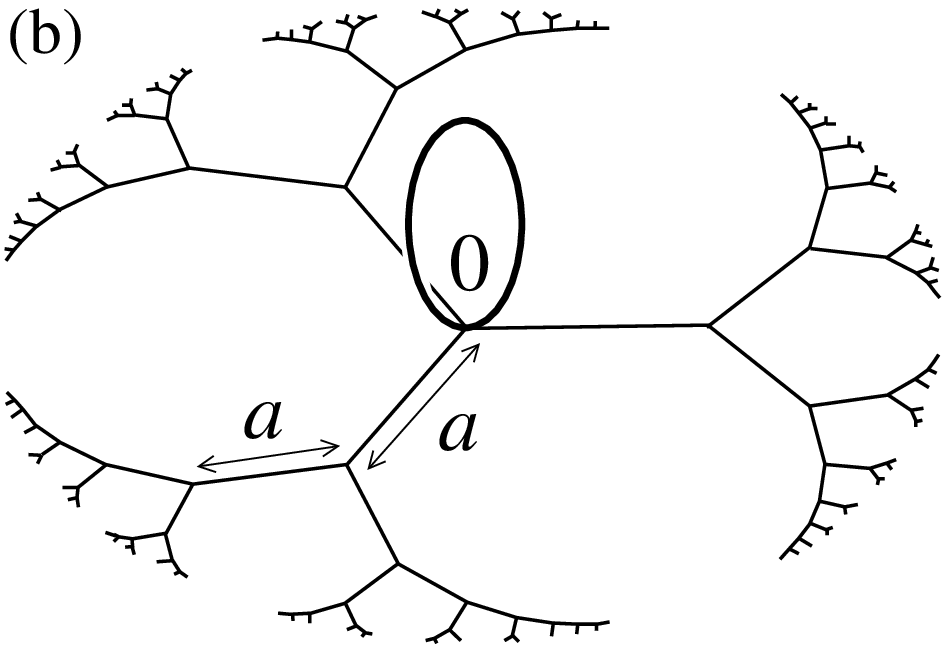}
\end{center}
\caption{\it A diffusive ring attached to {\rm (a)~:} an infinite square 
         lattice,
         {\rm (b)~:} an infinite Bethe lattice of coordinence $z=3$. 
         In both cases the bonds have equal lengths $a$.
         \label{fig:ringmesh}}
\end{figure}

The above examples show that the harmonics of the AAS oscillations of
a mesoscopic  ring  can change drastically due to the excursion of
Brownian trajectories in the arms connected to the ring. An
interesting question is how the harmonics are modified if the ring
is connected to a network more complex than a wire~? 
An important point is to know whether the Brownian
motion inside the network is recurrent or not
(a Brownian motion is said to be recurrent if it comes back to its starting
point with probability one after an infinite time).
The fact that the one-dimensional random walk is recurrent is
crucial to lead to the behaviour (\ref{harmT14b},\ref{Lapdis},\ref{anobeh}), 
as explained in section \ref{sec:comb}. 
In dimension $d>2$ the Brownian motion is known to be not recurrent
whereas the case of $d=2$ is marginal. Therefore we expect the dimension 2
to play a special role. This is one of the reasons why we consider
below the case of a ring connected to a square two-dimensional
network, as shown on the figure~\ref{fig:ringmesh}a. 
The formalism is introduced for an arbitrary network.

First we consider the network without the loop. It is 
characterized by a matrix 
whose determinant gives the spectral determinant $S_{\rm net}(\gamma)$.

If we now attach a ring at the vertex $0$ of the network, the new network 
is characterized by a matrix ${\cal M}={\cal M}_{\rm net}+\delta{\cal M}$
with
\be
\delta{\cal M}_{\alpha\beta} 
= A(\theta,\lambda_0)\,
\delta_{\alpha0}\delta_{\beta0}
\ee
where (appendix C of \cite{AkkComDesMonTex00})
\be 
A(\theta,\lambda_0) = 
\lambda_0+ 2 \sqrt{\gamma} \,
\frac{ \cosh\sqrt{\gamma}L - \cos\theta }{ \sinh\sqrt{\gamma}L }
\:.\ee
Since $\delta{\cal M}$ has only one nonzero element, $(\delta{\cal M})_{00}$,
we have
\be
\det{\cal M}=
\left[
  1+A(\theta,\lambda_0)\,({\cal M}_{\rm net}^{-1})_{00}
\right]\:\det{\cal M}_{\rm net}
\:.\ee
This shows that the spectral determinant of the network with the loop is 
now
\be
S^{(\lambda_0)}(\gamma) = \frac{\sinh\sqrt{\gamma}L}{\sqrt{\gamma}}
\left[
  1+A(\theta,\lambda_0)\,({\cal M}_{\rm net}^{-1})_{00}
\right]\:S_{\rm net}(\gamma)
\:.\ee
Using (\ref{localcoop}) we have
$
P_c(0,0)=
\frac{\D}{\D\lambda_0}
\ln[1+A(\theta,\lambda_0)\,({\cal M}_{\rm net}^{-1})_{00}]\big|_{\lambda_0=0}
$
that leads to the same result (\ref{Pc00},\ref{Wn00}) as 
when the ring is connected to an arm.
It is interesting to note that the matrix element is related to the Green's
function of the diffusion operator in the network, {\it i.e.} the cooperon at
the vertex $0$ in the absence of the ring~:
\be
P_c^{\rm net}(0,0) = \bra{0} \frac1{\gamma-\Delta} \ket{0} 
= ({\cal M}_{\rm net}^{-1})_{00}
\:.\ee
Therefore we can write the cooperon at the vertex $0$ in terms of the cooperon
of the network in the absence of the ring~:
\be
P_c(0,0) =\frac{1}{2\sqrt\gamma}
\frac{\sinh\sqrt\gamma L}
     {  \cosh\sqrt{\gamma}L
       +\frac12\frac{\sinh\sqrt{\gamma}L}{\sqrt{\gamma}P_c^{\rm net}(0,0)} 
       - \cos\theta} 
\:.\ee
Now the effective perimeter is related to the matrix ${\cal M}_{\rm net}$
characterizing the network in the absence of the loop~:
\be\label{leffnet}
\cosh\sqrt{\gamma} L_{\rm eff} 
=\cosh\sqrt{\gamma}L
+\frac12 
\frac{\sinh\sqrt{\gamma}L}{\sqrt{\gamma}\,({\cal M}_{\rm net}^{-1})_{00}}
\:,\ee
which is a generalization of the results (\ref{leffDir},\ref{leffNeu}) 
obtained for the ring attached to an arm.
We can easily check that if the network is simply  a wire 
with Dirichlet boundary at one end, we have 
$({\cal M}_{\rm net}^{-1})_{00}=1/(\sqrt\gamma \coth\sqrt\gamma b)$, 
leading to (\ref{leffDir}).
In the limit $\sqrt{\gamma}L\ll1$ ({\it i.e.} $L\ll L_\varphi$), 
eq.~(\ref{leffnet})
leads to 
\be
\sqrt{\gamma} L_{\rm eff} \simeq
\sqrt{\frac{L}{({\cal M}_{\rm net}^{-1})_{00}}}
=\sqrt{ \frac{L}{P_c^{\rm net}(0,0)} }
\:.\ee

\vspace{0.25cm}

\noindent{\bf The case of regular networks.--}
It is interesting to consider the specific case of a regular network.
In this case we can write~:
${\cal M}_{\rm net}=\frac{\sqrt{\gamma}}{\sinh\sqrt{\gamma}a}N$
where the matrix $N$ is 
$N_{\alpha\beta}=\delta_{\alpha\beta}\,z\cosh\sqrt{\gamma}a-a_{\alpha\beta}$. 
The matrix $a_{\alpha\beta}$ is the connectivity matrix (see 
appendix~\ref{sec:sd}).
We have introduced the coordinence $z$ of the network.
In this case  ${\cal M}_{\rm net}^{-1}$ is related to the (discrete) Green's 
function of the connectivity matrix $G(\alpha,\beta)=(N^{-1})_{\alpha\beta}$, 
defined by
$
\sum_\mu
(E\,\delta_{\alpha\mu} - a_{\alpha\mu}) G(\mu,\beta)=\delta_{\alpha\beta}
$,
for an ``energy'' $E=z\cosh\sqrt{\gamma}a$.
It follows that the effective length is related to the discrete Green's 
function of the network at the point where the ring is attached 
\be
\cosh\sqrt{\gamma} L_{\rm eff} =\cosh\sqrt{\gamma}L
+\frac12 \frac{\sinh\sqrt{\gamma}L}{\sinh\sqrt{\gamma}a\:G(0,0)}
\:.\ee
In the limit $\sqrt{\gamma}L\ll1$, we have
\be
\sqrt{\gamma} L_{\rm eff} \simeq
\sqrt{\frac{\sqrt{\gamma}L}{\sinh(\sqrt{\gamma}a)\:G(0,0)}}
\:.\ee

\subsection{Ring connected to a square network}

We now apply these results to the case of a square lattice. The problem left
is to estimate the discrete Green's function of the square network (figure
\ref{fig:ringmesh}a). We have
\be\label{fG00}
G(0,0) =
\frac12\int_{-\pi}^{+\pi}\frac{\D^2\vec k}{(2\pi)^2}
\frac1{2\cosh\sqrt{\gamma}a-\cos k_x-\cos k_y}
=\frac1{2\pi\cosh\sqrt{\gamma}a}
\: {\rm K}\!\left(\frac1{\cosh\sqrt{\gamma}a}\right) 
\ee 
where ${\rm K}(x)$
is the complete elliptic integral of first kind. 
We have a new characteristic time $\tau_a=a^2$ needed to explore the bond.
We now consider two limits of interest: 

\vspace{0.25cm}

\noindent$\bullet$ {\bf Intermediate time} 
\mathversion{bold} $\tau_L\ll t\ll\tau_a$ \mathversion{normal}
({\it i.e.} $L\ll L_\varphi\ll a$).--
In this limit the Brownian
trajectories starting from the ring can only explore a small
portion of the 4 arms to which it is connected. Eq.~(\ref{fG00}) gives
$ L_{\rm eff} \simeq\sqrt{4L}\gamma^{-1/4}$. This result corresponds
precisely to the case of a ring connected to $N_a=4$ long arms and we
have
$
\int_0^\infty\D t\,W_n(0,0;t)\,\EXP{-\gamma t}
\propto\exp-n\sqrt{4L}\gamma^{1/4}
$.

\medskip

\noindent$\bullet$ {\bf Long time}  
\mathversion{bold} $\tau_L,\tau_a\ll t$ \mathversion{normal}
({\it i.e.}  $L,a\ll L_\varphi$).-- In
this case the Brownian trajectories encircling
the flux $n$ times can explore the square network over  distances
much larger than $a$. Eq.~(\ref{fG00}) leads to
\be\label{leffsn}
\sqrt\gamma\, L_{\rm eff}\simeq\sqrt{\frac{2\pi L}{a\ln(4/\sqrt{\gamma}a)}}
\:.\ee 
The probability then
reads 
\be\label{fcsn}
\int_0^\infty\D t\,W_n(0,0;t)\,\EXP{-\gamma t} \simeq
\sqrt{\frac{La}{8\pi}\ln(4/\sqrt{\gamma}a)}\:
\exp-n\sqrt{\frac{2\pi L}{a\ln(4/\sqrt{\gamma}a)}} 
\:.\ee 
The Brownian trajectories encircling the ring can leak over long
distances in the square network. The effective diffusion in the
ring is even slowed down compare to the case of a ring connected
to arms. The number of windings behaves like~:
\be\label{ssn}
n_t\simeq\sqrt{\frac{a}{\pi L}\ln t} 
\:\ee
(the way to obtain the precise coefficient is explained in the 
appendix~\ref{sec:nt}).
An argument similar to the one of section \ref{sec:comb} can be developed
to obtain (\ref{ssn}) by different means. 
Since the diffusion is recurrent in the square network, this latter
acts as a trap in which the diffusive particle stays during a time 
$\tau$ and eventually come back inside the ring.
The distribution of the trapping time is given by the first return
probability in a square lattice, that is known to behave at large times 
like $P_1(\tau)\propto1/(\tau\ln^2\tau)$ \cite{BarTac93}
\footnote{
  In this work the authors considered the survival probability 
  $W_{\rm abs}(r,t)$ for a particle diffusing from $r$ in the 
  presence of an absorbing site at 0.
  The probability for the particle to reach the origin for the first 
  time is proportional to $-\partial_tW_{\rm abs}(r,t)$.
}.
It can be shown from this distribution that the winding number scales 
like  $n_t\propto\sqrt{\ln t}$ \cite{satya}.

The result (\ref{fcsn}) can be interpreted as the
amplitude of the $n$-th harmonic of the AAS oscillations of the conductivity~:
\be
\Delta\sigma^{(n)}\propto\exp-n\sqrt{\frac{2\pi L}{a\ln(4L_\varphi/a)}}
\hspace{0.5cm}\mbox{for } L,a\ll L_\varphi
\:.\ee

\subsection{Ring connected to networks of higher dimensions}

To emphasize the role of the dimension of the network, let us consider 
a $d$-dimensional hypercubic network. 
The Green's function reads
$
G(0,0)=\frac12\int_{\rm BZ}\frac{\D^d\vec k}{(2\pi)^d}\,
[d\cosh\sqrt\gamma a-\sum_{i=1}^d\cos k_i]^{-1}
$, 
where the integration is performed over the Brillouin zone.
This can be conveniently rewritten 
$G(0,0)=\frac12\int_0^\infty\D y\,[\EXP{-y\cosh\sqrt\gamma a}I_0(y)]^d$ 
where $I_0(y)$ is the modified Bessel function.
({\it i}) In dimension $d=1$ and $d=2$, the integral is dominated by the
neighbourhood of $\vec k\sim0$ (or the domain of large $y$) in the limit 
$\sqrt{\gamma}a\ll1$. We can write
$
G(0,0)\simeq\int_{\rm BZ}\frac{\D^d\vec k}{(2\pi)^d}\,
\frac1{\vec k^2+\gamma a^2d/2}
$
or 
$
G(0,0)\simeq
\frac12\int^\infty\D y\,\frac{1}{(2\pi y)^{d/2}}\EXP{-y\gamma a^2d/2}
$.
In dimension $d=1$ the integral diverges as 
$G(0,0)\sim1/a\sqrt\gamma$, which reflects the recurrence of the 1d
Brownian motion and leads to 
$\sqrt\gamma L_{\rm eff}\sim\gamma^{1/4}L^{1/2}$.
The dimension $d=2$ is the marginal case~: the integral diverges 
logarithmically $G(0,0)\sim\ln(1/a\sqrt\gamma)$ which still indicates
a recurrent Brownian motion and gives (\ref{leffsn}).
({\it ii}) In dimension $d>2$ the integral reaches a finite value 
$G(0,0)=\frac12\int_0^\infty\D y\,[\EXP{-y}I_0(y)]^d=\beta_d$ 
in the limit $\gamma\to0$. Note that $\beta_d\simeq1/(2d)$ for large 
dimensions, that coincides  with the result given below for a Bethe lattice
of coordinence $z=2d$.
The Brownian motion is not recurrent. 
Therefore $\sqrt\gamma L_{\rm eff}$ is independent on 
$\gamma$, which indicates that the winding number $n_t$ reaches a finite 
value in the infinite time limit.

\vspace{0.25cm}

\noindent{\bf Bethe lattice.--}
Let us consider the case of the Bethe lattice of coordinence $z$ 
(figure~\ref{fig:ringmesh}b), that  models a network with an infinite 
effective dimension.
The Green's function at coinciding point is \cite{MonTex96}
\be
G(0,0) = \frac1E F_0\left(\frac{z}{E}\right)
\hspace{0.25cm}\mbox{ with }\hspace{0.25cm}
F_0(x) = \frac{2(z-1)}{z-2+\sqrt{z^2 - 4(z-1)x^2}}
\:.\ee
We obtain the following behaviour for $t\gg\tau_L,\,\tau_a$
\be
\int_0^\infty\D t\,W_n(0,0;t)\,\EXP{-\gamma t} \propto
\exp-n\sqrt{\frac{z(z-2)}{z-1}\frac{L}{a}}
\:.\ee
This result is similar to the one obtained for a ring connected to 
an arm with absorption at the end (Dirichlet condition). In this latter
case the absorption at the end of the arm is responsible from the fact 
that the particle leaves the ring after a finite time.
In  the Bethe lattice, since the diffusion is not recurrent, 
the diffusive particle injected in the ring eventually get lost in the 
infinite lattice.
At large times the number of windings in the ring reaches the finite limit~:
\be
n_t\simeq\sqrt{\frac{2(z-1)}{z(z-2)}\frac{a}{L}}
\:.\ee
The higher the coordinence, the smaller the time spent by the particle 
in the ring is. The shorter $a$, the faster the particle feels the 
structure of the Bethe lattice and get lost in it.

\vspace{0.25cm}

All results are summarized in table~\ref{Latable}.

\begin{table}
\begin{center}
\begin{tabular}{|l|c|c|lc|}
\hline
Network  & Regime                & $ L_{\rm eff}(\gamma)$ &  winding
                                                  & $n^2_t$   \\\hline\hline
no       & $t\ll\tau_L$          & $L$           & normal
                                              & $2\exp-\frac{L^2}{4t}$\\[0.1cm]
         & $\tau_L\ll t$         & $L$&            normal 
                                                  & $2t/L^2$ \\[0.1cm]
\hline\hline
$N_a$ arms& $t\ll\tau_L,\,\tau_b$ & $L$           & normal
                                              & $2\exp-\frac{L^2}{4t}$\\[0.1cm]
         & $\tau_L\ll t\ll\tau_b$& $\sqrt{N_aLL_\varphi}$ & anomalous 
                                             & $2\sqrt{\pi t}/(N_aL)$ \\[0.1cm]
\hspace{0.45cm}
Neumann  &                       & $\sqrt{L(L+N_a b)}$& normal 
                                             & $2t/(L^2+N_abL)$ \\[-0.10cm]
\hspace{0.6cm}
Dirichlet&\raisebox{0.3cm}{$\tau_L,\,\tau_b\ll t$}& $L_\varphi\sqrt{N_a L/b}$
                                 & limited       & $2b/(N_a L)$ \\[0.1cm]
\hline\hline
square   & $\tau_L\ll t\ll\tau_a$& $\sqrt{4LL_\varphi}$ & anomalous 
                                                 & $\sqrt{\pi t}/(2L)$\\[0.1cm]
         & $\tau_L,\,\tau_a\ll t$     
         & $L_\varphi\sqrt{\frac{2\pi L}{\ln(4L_\varphi/a)}}$
         & anomalous & $\frac{a}{\pi L}\ln(t)$\\[0.1cm]
\hline\hline
$d$-hypercubic
         & $\tau_L\ll t\ll\tau_a$& $\sqrt{2dLL_\varphi}$ & anomalous 
                                                 & $\sqrt{\pi t}/(dL)$\\[0.1cm]
\hspace{1.2cm}$d>2$         
         & $\tau_L,\,\tau_a\ll t$     
         & $L_\varphi\sqrt{\frac1{\beta_d}\frac{L}{a}}$
         & limited &  $2\beta_d\frac{a}{L}$ \\[0.1cm]
\hline\hline
Bethe    & $\tau_L\ll t\ll\tau_a$& $\sqrt{zLL_\varphi}$ & anomalous 
                                                & $2\sqrt{\pi t}/(zL)$\\[0.1cm]
         & $\tau_L,\,\tau_a\ll t$     
         & $L_\varphi\sqrt{\frac{z(z-2)}{z-1}\frac{L}{a}}$
         & limited &  $\frac{2(z-1)}{z(z-2)}\frac{a}{L}$ \\[0.1cm]
\hline
\end{tabular}
\end{center}
\caption{\it 
         We summarize the results for the winding around the ring of 
         perimeter $L$ connected to a network.
         We recall that $\gamma=1/L_\varphi^2$ allows to probe the various 
         time regimes since it is the parameter conjugate to 
         time~:~$\gamma\sim1/t$.
         There are in general three characteristic times~:
         the time $\tau_L=L^2$ to diffuse around the ring,
         the time $\tau_a=a^2$ to explore one bond of the network
         and the time $\tau_b=b^2$ to reach the boundary of the network
         of linear size $b$.
         Except for the case of the wire, we have considered infinite
         networks with $b=\infty$.
         \label{Latable}}
\end{table}


\section{Conclusion}

We have studied the winding properties inside a ring connected to a network.
Our analysis was based on the introduction of the effective perimeter $ L_{\rm
eff}(\gamma)$ that probes the winding at time scale $t\sim1/\gamma$. We have
obtained this effective perimeter as a function of the matrix 
${\cal M}_{\rm net}$ describing the network that is connected to the ring. The
analysis of the effective perimeter immediatly gives the nature of the
winding (normal or anomalous) since the winding number scales with time as
$n_t\sim\sqrt{t}/ L_{\rm eff}(1/t)$.

\begin{figure}[!ht]
\begin{center}
\includegraphics[scale=0.5]{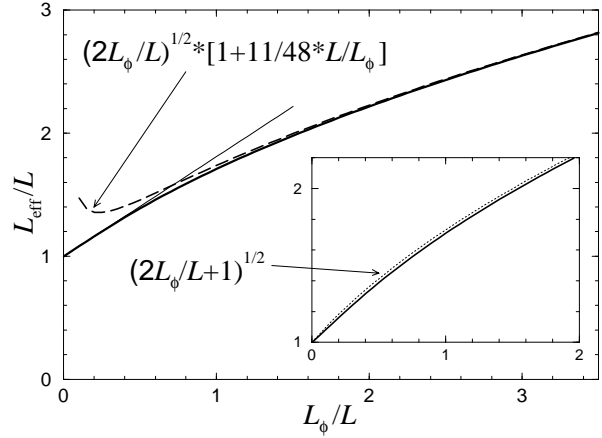}
\end{center}
\caption{\it The effective perimeter $L_{\rm eff}$ for the ring of 
         figure~\ref{fig:loop2} with $N_a=2$ 
         arms, as a function of the phase coherence 
         length. $L_{\rm eff}$ is given by 
         eq.~(\ref{leff2arms}). We have taken the limit $b\gg L_\varphi$.
         The dashed line is the expansion for large phase coherence length
         $L_{\rm eff}
         =\sqrt{2L_\varphi L}(1+\frac{11}{48}\frac{L}{L_\varphi}+\cdots)$.
         The thin line is the linear expansion 
         $L_{\rm eff}
         =L+L_\varphi\ln(9/4)+\cdots$ at small $L_\varphi$.
         In the inset, we compare the exact result with the approximation
         $\sqrt{L^2+2L_\varphi L}$ (dotted line).
          \label{fig:crossover}}
\end{figure}

Our study of winding properties was motivated by the physics of quantum 
transport through a ring.
We have emphasized the importance of taking  properly into account the
external wires connecting the ring when studying quantum transport.
Experimentally, the measurement of ratio of harmonics provides a direct
way to extract the phase coherence length and its temperature dependence 
(this has been used very recently on large square networks 
\cite{FerAngRowGueBouTexMonMai04}). 
In particular assuming an exponential behaviour $\exp-nL/L_\varphi$,
given by (\ref{AAS}), or the nonexponential $\exp-n\sqrt{2L/L_\varphi}$,  
given by (\ref{harmT14b}), leads to very different temperature dependences
of $L_\varphi(T)$.
The figure~\ref{fig:crossover} summarizes our main result~: we have plotted 
the effective length as a function of $L_\varphi$. The linear behaviour
at small $L_\varphi$, corresponds to (\ref{harmT14a}) and the square root 
behaviour at large $L_\varphi$, to (\ref{harmT14b}).
Since the crossover between the two regimes occurs on the range 
$L\sim L_\varphi$, it is useful to know the nature of this crossover.
We note that the effective perimeter for the ring with $N_a$ arms
can be well approximated by $L_{\rm eff}\simeq\sqrt{L^2+N_aL_\varphi L}$
which interpolates between $L$ and $\sqrt{N_aL_\varphi L}$ (see inset of
figure~\ref{fig:crossover}).
Therefore the harmonics decay approximatively as  
$\exp-n\sqrt{(L/L_\varphi)^2 + N_aL/L_\varphi}$. It is clear from this 
approximation that the crossover, that occurs for $L_\varphi\simeq L/N_a$, 
can be more easily reached for a large number of arms.
In a metallic ring, the phase coherence can reach several microns. 
Therefore, the regime $L_\varphi>L/N_a$ seems to be
attainable experimentally.


The phase coherence length $L_\varphi$ has been introduced as an effective 
parameter $\gamma=1/L_\varphi^2$ in the cooperon
$P_c(x,x')=\bra{x}\frac1{\gamma-\Delta}\ket{x'}$
in order to describe phase breaking mechanisms. 
However, the effect of electron-electron interactions, that dominates 
at low temperature, is not well described by such an effective parameter
\cite{AltAroKhm82}.
It has been shown recently  that the behaviour of the AAS 
harmonics in a ring behaves in this case as $\exp-nT^{1/2}L^{3/2}$ in the 
limit $L_\varphi\ll L$ , where $T$ is the temperature
\cite{LudMir04}. 
However this behaviour has not been observed 
in the recent experiments \cite{FerAngRowGueBouTexMonMai04} on large 
square networks, in which the condition $L_\varphi\ll L$ is not well 
fulfilled. A mechanism similar to the one discussed in our paper could
explain the discrepancy between the theory and the 
experiment~:
for $L_\varphi\gtrsim L$ the excursion of the Brownian curves of finite
winding in the surrounding part of the network increases the effective
perimeter of these paths and modifies the dependence of the harmonics in
$L/L_\varphi$.


\section*{Acknowledgements}

C.~T. acknowledges stimulating discussions with Alain Comtet and Jean Desbois.
We are grateful to Satya Majumdar for his enlighting remarks and having pointed
to our attention Ref.~\cite{BarTac93}.


\begin{appendix}

\section{Spectral determinant\label{sec:sd}}

A particularly convenient way to describe the spectrum $\{E_n\}$ of the 
Laplace operator $\Delta$ on a network of one-dimensional wires is to 
consider the spectral determinant 
$S(\gamma)=\det(\gamma-\Delta)=\prod_n(\gamma+E_n)$ where $\gamma$ 
is the spectral parameter. If the loops of the network are pierced by 
magnetic fluxes the derivative is replaced by the covariant derivative~:
$\Delta\to \Dc_x^2$ with $ \Dc_x=\frac{\D}{\D x}-\I A(x)$.

Let us consider a network of $B$ wires connected at $V$ vertices. These latters
are labelled with greek indices $\alpha$, $\beta$,\ldots
First we introduce the connectivity matrix $a_{\alpha\beta}$ characterizing
the topology of the network~: $a_{\alpha\beta}=1$ is $\alpha$ are $\beta$
connected by a bond (we denote $(\alpha\beta)$ the bond and
$l_{\alpha\beta}$ its length). $a_{\alpha\beta}=0$ otherwise.
If we define a scalar function $\psi(x)$ on the network, boundary
conditions at the vertices must be specified.
At the vertex $\alpha$ we choose ({\it i}) continuity of the
function, that is all the components $\psi_{\alpha\beta}(x)$ of
the function along the wires $\alpha\beta$ issuing from $\alpha$
tend to the same limit as the coordinate reaches the vertex.
({\it ii})
$
\sum_\beta a_{\alpha\beta}\Dc_x\psi_{\alpha\beta}(\alpha)=
\lambda_\alpha\psi(\alpha)
$
where the connectivity matrix in the
sum constrains it to run over the neighbouring vertices of $\alpha$.
Therefore the sum runs over all wires issuing from the vertex. 
The real parameter $\lambda_\alpha$ allows to describe several 
boundary condition: $\lambda_\alpha=\infty$ forces the function to 
vanish, $\psi(\alpha)=0$, and corresponds to Dirichlet boundary 
condition that describes the connection to a reservoir that absorbs 
particles.
$\lambda_\alpha=0$ corresponds to Neumann condition, which ensures 
conservation of the probability current and describes an internal vertex.

It was shown in \cite{PasMon99} that the spectral determinant is~:
\be
S(\gamma) =
\prod_{(\alpha\beta)}\frac{\sinh\sqrt\gamma l_{\alpha\beta}}{\sqrt\gamma}\:
\det{\cal M}(\gamma)
\ee
where the product runs over all bonds of the network.
${\cal M}$ is a $V\times V$ matrix defined as~:
\be\label{defM}
{\cal M}_\ab=\delta_\ab
\left(\lambda_\alpha + \sqrt\gamma\sum_\mu a_{\alpha\mu}
      \coth(\sqrt{\gamma}l_{\alpha\mu})\right)
-a_\ab\frac{\sqrt\gamma\:\EXP{-\I\theta_\ab}}{\sinh(\sqrt{\gamma}l_\ab)}
\:,\ee
where the connectivity matrix constrains the sum to run over all vertices 
$\mu$ connected to $\alpha$.
We have included the magnetic fluxes ($\theta_\ab$ is the flux along the 
wire).
For a more detailed presentation, see \cite{AkkComDesMonTex00}.

\section{Time dependence of the winding number\label{sec:nt}}

We explain how to compute efficiently the time dependence of the 
winding number around a ring connected to a network. 
For this purpose we compute $\smean{n^2}_t$, where the average is
taken for all close trajectories starting from a specified point and
coming back to it after a time $t$~: 
\be
\smean{n^2}_t = \frac{\sum_n n^2\,W_n(x,x;t)}{\sum_nW_n(x,x;t)}
\:\ee
($\smean{n}_t=0$ follows from the symmetry $P_c|_\theta=P_c|_{-\theta}$).

To go further we consider the case where the initial condition $x$ is
the vertex $0$ where the ring is attached to the network. Then the cooperon
has the structure (\ref{Pc00}).
Let us define the quantity
\be
\Omega_m(\gamma) = \left(\frac1\I\frac{\D}{\D\theta}\right)^m 
P_c(0,0)\big|_{\theta=0}
\:.\ee
From (\ref{Wn}) it follows that 
$\Omega_m(\gamma)=\int_0^\infty\D t\,\EXP{-\gamma t}\sum_n n^mW_n(0,0;t)$
which is related to the $m$-th moment of the winding number.
We have 
\be
\smean{n^2}_t= 
\frac{{\cal L}^{-1}\left[\Omega_2(\gamma)\right]}
     {{\cal L}^{-1}\left[\Omega_0(\gamma)\right]}
\ee
where ${\cal L}^{-1}\left[\cdots\right]$ designates the inverse Laplace 
transform.
From (\ref{Pc00}) we can obtain the two general expressions~:
\bea
\Omega_0(\gamma)&=&\frac{1}{4\sqrt\gamma}\,
\frac{\sinh\sqrt\gamma L}
     {\left(\sinh\frac{\sqrt\gamma  L_{\rm eff}}{2}\right)^2}\\
\Omega_2(\gamma)&=&\frac{1}{8\sqrt\gamma}\,
\frac{\sinh\sqrt\gamma L}
     {\left(\sinh\frac{\sqrt\gamma  L_{\rm eff}}{2}\right)^4}
\:.\eea

\vspace{0.25cm}

\noindent{\it Example 1~: the isolated ring.--}
The effective perimeter is equal to the perimeter in this case~: 
$ L_{\rm eff}=L$. We have
$\Omega_0(\gamma)=\frac{1}{2\sqrt\gamma}\coth(\sqrt\gamma L/2)$
and 
$
\Omega_2(\gamma)
=\frac{1}{4\sqrt\gamma}\coth(\sqrt\gamma L/2)/\sinh^2(\sqrt\gamma L/2)
$.
We consider two regimes~: \\
$\bullet$ Short times $t\ll\tau_L$~:
We have $\Omega_0(\gamma)\simeq\frac{1}{2\sqrt\gamma}$ and 
$\Omega_2(\gamma)\simeq\frac{1}{\sqrt\gamma}\EXP{-\sqrt\gamma L}$ which gives 
after Laplace transform
$\smean{n^2}_t \simeq 2\exp-\frac{L^2}{4t}$. This result was obvious from the 
expression (\ref{normaldiff}).\\
$\bullet$ Long times $t\gg \tau_L$~:
We obtain in this case $\Omega_0(\gamma)\simeq\frac{1}{\gamma L}$ 
and $\Omega_2(\gamma)\simeq\frac{2}{\gamma^2L^3}$, 
which leads to $\smean{n^2}_t \simeq 2t/L^2$. This is the normal diffusion.

\vspace{0.25cm}

\noindent{\it Example 2~: the ring connected to a square network.--}
We  consider the long time limit $t\gg \tau_L,\tau_a$ to demonstrate 
(\ref{ssn}). 
The effective perimeter
is given by (\ref{leffsn}). We find 
$\Omega_0(\gamma)\simeq\frac{a}{4\pi}\ln(16/a^2\gamma)$ and 
$\Omega_2(\gamma)\simeq\frac{a^2}{8\pi^2L}\ln^2(16/a^2\gamma)$.
Using the fact that in the limit $\gamma\to0$ we have 
${\cal L}^{-1}\left[\ln1/\gamma\right]\simeq1/t$
and ${\cal L}^{-1}\left[\ln^21/\gamma\right]\simeq2\ln(t)/t$,
we eventually find
$\smean{n^2}_t \simeq \frac{a}{\pi L}\ln t$.

\end{appendix}




\end{document}